\documentclass[ALICE,manyauthors]{cernphprep}
\usepackage[comma,square,numbers,sort&compress]{natbib}
\usepackage{hyperref} 
\usepackage{lineno}
\usepackage{color}
\usepackage{graphicx}
\usepackage{epstopdf}
\usepackage{amssymb}
\usepackage{xspace}
\usepackage{amsmath}
\usepackage{upgreek}
\usepackage{hyperref}
\usepackage{doi}
\usepackage{multirow}
\usepackage{makecell}
\usepackage{booktabs}
\usepackage[T1]{fontenc}
\usepackage{orcidlink}

\DeclareRobustCommand{\unit}[2][]{%
        \begingroup%
                \def\0{#1}%
                \expandafter%
        \endgroup%
        \ifx\0\@empty%
                \ensuremath{\mathrm{#2}}%
        \else%
                \ensuremath{#1\,\mathrm{#2}}%
        \fi%
        }
\DeclareRobustCommand{\unitfrac}[3][]{%
        \begingroup%
                \def\0{#1}%
                \expandafter%
        \endgroup%
        \ifx\0\@empty%
                \raisebox{0.98ex}{\ensuremath{\mathrm{\scriptstyle#2}}}%
                \nobreak\hspace{-0.15em}\ensuremath{/}\nobreak\hspace{-0.12em}%
                \raisebox{-0.58ex}{\ensuremath{\mathrm{\scriptstyle#3}}}%
        \else
                \ensuremath{#1}\,%
                \raisebox{0.98ex}{\ensuremath{\mathrm{\scriptstyle#2}}}%
                \nobreak\hspace{-0.15em}\ensuremath{/}\nobreak\hspace{-0.12em}%
                \raisebox{-0.58ex}{\ensuremath{\mathrm{\scriptstyle#3}}}%
        \fi%
}

\topfigrule
\newcommand {\meanpT}    {\ensuremath{\langle p_{\mathrm{T}} \kern-0.1em\rangle}\xspace}
\newcommand {\mean}[1]   {\ensuremath{\langle #1 \kern-0.1em\rangle}\xspace}

\newcommand {\ee}        {\mbox{$\mathrm {e^+e^-}$}\xspace}
\newcommand {\ep}        {\mbox{$\mathrm {e\kern-0.05em p}$}\xspace}
\newcommand {\pp}        {\mbox{$\mathrm {p\kern-0.05em p}$}\xspace}

\newcommand {\ppBoldMath} {\mbox{$\mathrm { \mathbf p\kern-0.05em \mathbf p }$}\xspace}

\newcommand {\MeanNpart} {\mbox{\ensuremath{< \kern-0.15em N_{part} \kern-0.15em >}}}

\newcommand {\mass}         {\mbox\mathrm{MeV$\kern-0.15em /\kern-0.12em c^2$}}
\newcommand {\tev}          {\ensuremath{\mathrm{TeV}}\xspace}
\newcommand {\gev}          {\ensuremath{\mathrm{GeV}}\xspace}
\newcommand {\mev}          {\ensuremath{\mathrm{MeV}}\xspace}

\newcommand {\mmom}         {\ensuremath{\mathrm{MeV\kern-0.15em /\kern-0.12em c}}\xspace}
\newcommand {\gmom}         {\ensuremath{\mathrm{GeV\kern-0.15em /\kern-0.12em c}}\xspace}
\newcommand {\mmass}        {\ensuremath{\mathrm{MeV\kern-0.15em /\kern-0.12em c^2}}\xspace}
\newcommand {\gmass}        {\ensuremath{\mathrm{GeV\kern-0.15em /\kern-0.12em c^2}}\xspace}

\newcommand {\dg}           {\ensuremath{\kern+0.1em ^\circ}\xspace}

\newcommand{\pt}            {\ensuremath{p_\mathrm{T}}\xspace}

\newcommand{\Jpsi}          {\ensuremath{\mathrm{J}/\uppsi}\xspace}

\newcommand{\DstartoDpi}    {\ensuremath{\mathrm{D}^{*+} \to \mathrm{D}^0 (\to \mathrm{K}^-\uppi^+) \uppi^+}\xspace}

\newcommand{\Dzero}         {\ensuremath{\mathrm{D^0}}\xspace}

\newcommand{\Dstar}         {\ensuremath{\mathrm{D^{*+}}}\xspace}

\newcommand{\Lb}            {\ensuremath{\mathrm{{\Lambda_b^0}}}\xspace}

\newcommand{\ZZero}         {\ensuremath{\mathrm{Z}^0}\xspace}

\newcommand{\GeV}           {\gev}
\newcommand{\MeV}           {\mev}

\newcommand{\TeV}           {\tev}
\newcommand{\Lint}          {\ensuremath{\mathcal{L}_\mathrm{int}}\xspace}

\newcommand{\mum}           {\ensuremath{\mathrm{\mu m}}\xspace}

\newcommand{\fprompt}       {\ensuremath{f_\mathrm{prompt}}\xspace}
\newcommand{\fnonprompt}    {\ensuremath{f_\mathrm{non\text{-}prompt}}\xspace}
\newcommand{\AccEff}        {\ensuremath{\mathrm{Acc}\times\epsilon}\xspace}
\newcommand{\effP}[1]       {\ensuremath{(\mathrm{Acc}\times\epsilon)_{\mathrm{prompt}, #1}}\xspace}
\newcommand{\effNP}[1]      {\ensuremath{(\mathrm{Acc}\times\epsilon)_{\mathrm{non\text{-}prompt},#1}}\xspace}
\newcommand{\rawY}[1]       {\ensuremath{Y_#1}\xspace}

\newcommand{\HbtoDstar}     {\ensuremath{\mathrm{H_b\to\Dstar+X}}\xspace}

\newcommand{\rzz}           {\ensuremath{\rho_{00}}\xspace}
\newcommand{\cost}          {\ensuremath{\cos{\theta^*}}\xspace}
\newcommand{\Np}               {\ensuremath{N_\mathrm{prompt}}\xspace}
\newcommand{\Nnp}              {\ensuremath{N_\mathrm{non\text{-}prompt}}\xspace}

\newcommand{\s}               {\ensuremath{\sqrt{s}}\xspace}

\newcommand{\dstar}           {\ensuremath{\mathrm{D^{*+}}}\xspace}

\newcommand{\pythia}          {\textsc{Pythia~8}\xspace}
\newcommand{\herwig}          {\textsc{Herwig~7}\xspace}
\newcommand{\pythiaprec}      {\textsc{Pythia~8.243}\xspace}

\newcommand{\evtgen}          {\textsc{EvtGen}\xspace}
\newcommand{\geant}           {\textsc{Geant~3}\xspace}

\usepackage{float} 

\begin{document}%
\begin{titlepage}

\PHyear{2022}
\PHnumber{283}      % required, will be obtained from PH
\PHdate{12 December}  % required, will be obtained from PH %
  
  %%% Put your own title + short title here:

\title{First measurement of prompt and non-prompt $\pmb{\dstar}$ vector meson spin alignment in pp collisions at $\pmb{\s = 13}$ TeV}   
\ShortTitle{Spin alignment of \dstar vector mesons in pp collisions at $\s = 13$ TeV}
% appears on right page headers  
%%% Do not change the next lines
\Collaboration{ALICE Collaboration\thanks{See Appendix~\ref{app:collab} for the list of Collaboration members}}
\ShortAuthor{ALICE Collaboration} 
\begin{abstract}
  This letter reports the first measurement of spin alignment, with respect to the helicity axis, for \dstar vector mesons and their charge conjugates from charm-quark hadronisation (prompt) and from beauty-meson decays (non-prompt) in hadron collisions. The measurements were performed at midrapidity ($|y|<0.8$) as a function of transverse momentum (\pt) in proton–proton (pp) collisions collected by ALICE at the centre-of-mass energy $\s=13~\TeV$.
  The diagonal spin density matrix element \rzz of \Dstar mesons was measured from the angular distribution of the \DstartoDpi decay products, in the \dstar rest frame, with respect to the \dstar momentum direction in the pp centre of mass frame.
  The \rzz~value for prompt \dstar mesons is consistent with $1/3$, which implies no spin alignment. However, for non-prompt \dstar mesons an evidence of \rzz larger than $1/3$ is found. The measured value of the spin density element is $\rzz=0.455\pm0.022(\text{stat.})\pm0.035(\text{syst.})$ in the $5<\pt<20~\GeV/c$ interval, which is consistent with a \pythia Monte Carlo simulation coupled with the \evtgen package, which implements the helicity conservation in the decay of \dstar meson from beauty mesons. 
  In non-central heavy-ion collisions, the spin of the \Dstar mesons may be globally aligned with the direction of the initial angular momentum and magnetic field. Based on the results for pp collisions reported in this letter it is shown that alignment of non-prompt \Dstar mesons due to the helicity conservation coupled to the collective anisotropic expansion may mimic the signal of global spin alignment in heavy-ion collisions.
\end{abstract}
\end{titlepage}
\setcounter{page}{2}

\section{Introduction}
The production of hadrons containing heavy quarks, i.e.~charm and beauty, has been extensively studied in both lepton and hadron collisions, improving significantly the understanding of the hadronisation mechanism~\cite{Gladilin:2014tba,ALICE:2021dhb,LHCb:2019fns,HFLAV:2019otj,ALICE:2021npz,ALICE:2022cop,LHCb:2022syj,LHCb:2021qbv}. However, several aspects of the transition of the heavy quark to the final-state hadron are not yet settled. One of them regards the spin properties of produced particles in the quark hadronisation. 
In the limit of very large heavy-quark mass ($m_\mathrm{Q}\rightarrow\infty$), the polarisation of heavy-flavour baryons is assumed to be that carried by the heavy valence quark~\cite{Falk:1993rf}. 
This assumption was studied at LEP in \ee collisions where the polarisation of \Lb baryons was used to probe the electroweak structure of the \ZZero boson production and decays~\cite{Korner:1993dy,Mannel:1991bs,ALEPH:1995aqx,DELPHI:1999hkl,OPAL:1998wmk}. In hadronic collisions, the polarisation of \Lb baryons can arise during the hadronisation process if beauty quarks are produced with a transverse polarisation, as predicted by the heavy-quark effective theory~\cite{Mannel:1991bs,Falk:1993rf}. At this time, one measurement, performed by the LHCb Collaboration, found no significant polarisation~\cite{LHCb:2013hzx}.
For vector mesons with quantum numbers $J^P=1^-$, both in lepton and hadron collisions, at least part of the polarisation should arise from the hadronisation process~\cite{Falk:1993rf}. Models based on statistical quantum mechanics considerations~\cite{Donoghue:1978yb,Bigi:1977}, heavy-quark effective theory~\cite{Falk:1993rf}, or inspired by quantum chromodynamics (QCD)~\cite{Augustin:867018} provide very different predictions, from no polarisation, to very large longitudinal or transverse polarisation. This aspect of the hadronisation is typically not accounted for in QCD-inspired Monte Carlo (MC) generators based on the Lund string model~\cite{Andersson:1983ia}, such as \pythia~\cite{Sjostrand:2006za,Sjostrand:2014zea}, or the clustering model~\cite{Kersevan:2004yg}, as implemented in \herwig~\cite{Bellm:2015jjp}. Measurements of excited B vector meson states in \ZZero decays at LEP did not show any significant polarisation~\cite{DELPHI:1995hfn,ALEPH:1995ikc}. For charm vector mesons (\Dstar) a longitudinal polarisation was observed by OPAL~\cite{OPAL:1997nwj}. However, previous measurements in \ee collisions indicated no significant polarisation for \Dstar mesons~\cite{TPCTwo-Gamma:1991mvk,HRS:1987vlf,CLEO:1991bnf}.

The polarisation of vector mesons containing heavy quarks in heavy-ion collisions is of special interest. 
The initial stages of such collisions, with non-zero impact parameter, are expected to be characterised by a large orbital angular momentum~\cite{Becattini:2007sr} and a strong magnetic field~\cite{Kharzeev:2007jp,Skokov:2009qp,Deng:2012pc}. These initial conditions could influence the produced colour-deconfined matter (called quark--gluon plasma~\cite{Braun-Munzinger:2015hba,ALICE:2022wpn}).
The directions of the angular momentum and the magnetic field in such collisions are perpendicular to the reaction plane (subtended by the beam axis and impact parameter)~\cite{Liang:2004ph}. Theoretical calculations at LHC energies predict the values of the angular momentum to be of the order of $10^{7}$ $\hslash$~\cite{Becattini:2007sr} and the magnetic field to be about $10^{16}$ T~\cite{Christakoglou:2021nhe,Skokov:2009qp,Fraga:2008qn,Deng:2012pc,Kharzeev:2007jp}. While the angular momentum, a conserved quantity in strong interactions, is expected to affect the whole evolution of the collision, the magnetic field is transient in nature and its strength falls steeply with time. Heavy quarks are produced at the initial stages of heavy-ion collisions, in a time scale shorter than the quark--gluon plasma formation time~\cite{Andronic:2015wma} and are sensitive to the large initial magnetic field as well as the angular momentum. In the presence of these initial conditions, charm quarks can be polarised. The quark polarisation is expected to be further transferred to the final-state hadrons during hadronisation and it is predicted to be different in the case of hadronisation in vacuum, or recombination with light quarks from the deconfined medium~\cite{Liang:2004ph,Liang:2004xn,Liang:2007ma,Yang:2017sdk}. Measurements in heavy-ion collisions indicate that the recombination mechanism is important to describe the production and angular anisotropies of charm hadrons~\cite{ALICE:2021rxa,CMS:2017qjw,ALICE:2021kfc,STAR:2021tte,ALICE:2021bib,CMS:2019uws,STAR:2019ank,ALICE:2020iug,CMS:2020bnz}. If observed, a non-zero polarisation of \Dstar mesons will be a manifestation of the angular momentum and the magnetic field created in heavy-ion collisions. The study of \Dstar mesons would provide additional information, crucial for the interpretation of the recently measured \Jpsi polarisation in Pb--Pb collisions by the ALICE Collaboration~\cite{ALICE:2022sli,ALICE:2020iev}. In fact, the polarisation of \Dstar and \Jpsi mesons have different contributions from the angular momentum and the magnetic field, given the valence light quark present in the \Dstar meson, which is expected to be less sensitive than the charm quark to the magnetic field. 
However, the interpretation of experimental data for heavy-ion collisions requires a baseline study in elementary proton--proton (pp) collisions, where such initial conditions are not expected to form. Moreover, in this study, one of the major sources of background is the feed-down contribution of \Dstar mesons originating from decays of scalar mesons containing a beauty quark, which are observed to be longitudinally polarised by the Belle and LHCb Collaborations in \ee and pp collisions, respectively~\cite{Belle:2019ewo,LHCb:2021sqa}. This is a direct consequence of the helicity conservation and the Vector--Axial (V$-$A) nature of the weak interaction involved in the $\mathrm{b}\to\mathrm{c}$ decay, which leads to left-handed fermion couplings in interactions with W bosons~\cite{Suzuki:2001za, BaBar:2008lan}. It is therefore important to separate contributions to the polarisation from promptly produced mesons and from those originating in beauty-hadron decays. This was not possible in the case of the \Jpsi meson in Pb--Pb collisions with LHC Run 2 data, for which only the inclusive polarisation was used as baseline~\cite{ALICE:2020iev}.

The polarisation of \Dstar vector mesons cannot be probed directly as \Dstar mesons are measured through the parity conserving strong decay \DstartoDpi. However, experimentally, the spin alignment along the quantisation direction can be studied by measuring the diagonal spin density matrix element (\rzz) of the vector meson. The \rzz reflects the probability of finding a vector meson in the state with spin zero out of possible states with spin projection $-1$, 0, and 1~\cite{Fano:1957zz}. If the spin of particles is homogeneously distributed, all spin states are expected to be
equally probable, thereby yielding $\rzz = 1/3$. The spin alignment is studied by measuring the angular distribution of the decay products of vector mesons. Their decay products are measured with respect to a quantisation axis, which is the vector meson momentum direction (helicity axis) in pp collisions and the perpendicular direction of the reaction plane in heavy-ion collisions. In the latter, these two axes are further correlated through the anisotropic collective expansion which is quantified by coefficients in a Fourier decomposition of the azimuthal-angle distribution of final-state particle momenta, with the second-harmonic coefficient $v_2$, called elliptic flow~\cite{Snellings:2011sz}, being the dominant one~\cite{ALICE:2020iug}. 
The angular distribution of the \Dstar decay products is expressed as
\begin{equation}
\frac{\mathrm{d}N}{\mathrm{d}\cos{\theta^{*}}} \propto [ 1 - \rzz + (3\rzz - 1 )\cos^{2}{\theta^{*}} ],
\label{eqn1}
\end{equation}
where $\theta^{*}$ is the angle between the momentum of either the \Dzero meson or the pion in the rest frame of the vector meson with respect to the quantisation axis~\cite{Schilling:1969um}.

In this letter, the first measurement of the spin alignment of \Dstar mesons from charm-quark hadronisation (prompt) and from beauty-hadron decays (non-prompt) at midrapidity is presented as a function of transverse momentum (\pt) in pp collisions at a centre-of-mass energy of $\s = 13~\TeV$. Based on these measurements, quantitative predictions for the spin distributions of non-prompt \Dstar meson for a future measurement of the \Dstar-meson spin alignment in heavy-ion collisions are provided.

\section{Experimental apparatus and data analysis}
\label{sec:analysis}
The analysis was performed using about $1.8\times10^9$ minimum-bias and about $10^9$ high-multiplicity triggered pp collisions at $\s = 13~\TeV$ recorded with the ALICE apparatus~\cite{ALICE:2014sbx,ALICE:2008ngc}, corresponding to an integrated luminosity of $\Lint\approx32$  $\mathrm{nb}^{-1}$ and $\Lint\approx7.7~\mathrm{pb}^{-1}$, respectively. The minimum-bias trigger required coincident signals in the two V0 detectors, which are two scintillator arrays covering the pseudorapidity intervals $-3.7 < \eta < -1.7$ and $2.8 < \eta < 5.1$~\cite{ALICE:2013axi}. The high-multiplicity trigger selected the 0.17\% highest-multiplicity events out of all inelastic collisions with at least one charged particle in the pseudorapidity range $|\eta|<1$, relying on the signal amplitudes of the V0 detectors. Events were further selected offline to remove machine-induced backgrounds~\cite{ALICE:2014sbx}. They were required to have a reconstructed collision vertex located within $\pm10~\mathrm{cm}$ from the centre of the detector along the beam-line direction to maintain a uniform acceptance. Events with multiple reconstructed primary vertices (pileup) were rejected. Primary vertices were reconstructed from track segments measured with the two innermost layers of the Inner Tracking System (ITS), which consists of six cylindrical layers of silicon detectors~\cite{ALICE:2010tia}. The remaining undetected pileup was studied and found to be negligible for this analysis.

The \Dstar mesons and their charge conjugates were measured at midrapidity ($|y|<0.8$) via the \DstartoDpi decay channel, with branching ratio $\mathrm{BR} = (2.67\pm0.03)\%$~\cite{ParticleDataGroup:2022pth}. The measurement of \Dstar mesons was based on charged-particle tracks reconstructed with the Time Projection Chamber (TPC)~\cite{Alme:2010ke} and ITS detectors. The \Dzero decay candidates were defined combining pairs of tracks with the expected charge combinations. Each track from the decay was required to have $|\eta| < 0.8$, $\pt > 0.3~\gev/c$, at least 70 (out of 159) associated space points in the TPC, and a minimum of two hits in the ITS, with at least one in either of the two innermost layers to ensure a good pointing resolution. The \Dstar-meson candidates were reconstructed by combining \Dzero candidates with low-\pt tracks (referred to as soft pions) having $|\eta| < 0.8$, $\pt > 50~\mev/c$, and at least two hits in the ITS. The analysis was based on the reconstruction of decay-vertex topologies of \Dzero mesons displaced from the interaction vertex. In particular, the proper decay length of \Dzero mesons of $c\tau \approx 123~\mum$ and that of beauty hadrons of $c\tau \approx 500~\mum$ were exploited to resolve the \Dzero-meson decay vertices. In order to reduce the large combinatorial background and to separate the contribution of prompt and non-prompt \Dstar mesons, a multiclass classification algorithm based on Boosted Decision Trees (BDTs) was used~\cite{xgboost,hipe4ml}. The candidate information used by the BDT algorithm to distinguish among prompt and non-prompt \Dstar mesons and background candidates was based on i) the distance between the reconstructed \Dzero-meson decay vertex and the primary vertex, ii) the \Dzero-meson and soft-pion distance of closest approach to the interaction vertex, iii) the cosine of the pointing angle between the \Dzero-meson candidate line of flight and its reconstructed momentum vector, and iv) the particle identification (PID) information of the decay tracks. The PID information was provided by the specific energy loss and the flight time of particles measured with the TPC and Time Of Flight (TOF)~\cite{Akindinov:2013tea} detectors, respectively. Signal samples of prompt and non-prompt \Dstar mesons for the BDT training were obtained from MC simulations based on the \pythiaprec event generator~\cite{Sjostrand:2006za, Sjostrand:2014zea} with the Monash tune~\cite{Skands:2014pea} and \geant package~\cite{GEANT} for the propagation of the generated particles through the detector. The background samples were obtained in data from the sideband region of the invariant-mass distribution $\Delta M = M(\mathrm{K}\uppi\uppi) - M(\mathrm{K}\uppi)$. The sideband region was chosen as the invariant-mass interval $\Delta M > 150~\MeV/c^{2}$, where no \Dstar signal is present. Independent BDTs were trained for the high-multiplicity and minimum-bias samples in three different \pt intervals, i.e. $5<\pt<7~\GeV/c$, $7<\pt<10~\GeV/c$, and $10<\pt<20~\GeV/c$. Subsequently, they were applied to the real data sample in which the type of candidate is unknown. The BDT outputs are related to the candidate probability to be a prompt \Dstar meson, non-prompt \Dstar meson, or combinatorial background.

The selection based on the probability to be background was optimised to yield a good statistical significance and purity of the signal. By selecting high probability to be prompt or non-prompt \Dstar mesons, the samples of candidates were further separated to the charm-enhanced (significant contribution of prompt \Dstar mesons) and the beauty-enhanced (significant contribution of non-prompt \Dstar mesons) samples. The threshold values were tuned as a function of \pt using MC simulations. The fractions of prompt (\fprompt) and non-prompt (\fnonprompt) \Dstar mesons in each sample were evaluated with a data-driven method based on the sampling of the raw yield $Y_\mathrm{i}$ at different values of the BDT-output score related to the probability of being a non-prompt \Dstar meson. These fractions can be computed by solving a system of equations that relate raw yields to the corrected yields of prompt (\Np) and non-prompt (\Nnp) \Dstar mesons by the product of the geometrical acceptance and the reconstruction and selection (including BDT) efficiency factors (\AccEff) for prompt and non-prompt \Dstar mesons. Each equation, obtained for a set of BDT selections $i$, can be expressed as
\begin{equation}
   \effP{i}\times \Np +  \effNP{i}\times \Nnp - \rawY{i} = \delta_\mathrm{i},
\label{eq:eq_set}
\end{equation}
where $\delta_\mathrm{i}$ is the residual originating from the uncertainties on \rawY{i}, \effNP{i}, and \effP{i}. To define the sets of selections, the BDT-output response related to the probability of a candidate to be a non-prompt \Dstar meson was sampled monotonically, to obtain $n$ selections ordered in such a way that the $i^\mathrm{th}$ selected sample was completely included in the $(i-1)^\mathrm{th}$ one. The system of equation is then solved via a $\chi^2$ minimisation, which provides as output the corrected yields of prompt (\Np) and non-prompt (\Nnp) \Dstar mesons and their covariance matrix. A detailed description of this method is presented in Refs.~\cite{ALICE:2021mgk,ALICE:2022enj,ALICE:2022tji}.
For each \pt interval, the raw yields of \Dstar mesons were extracted with a fit to the distribution of the invariant mass $\Delta M$. 

\begin{figure}[!t]
\begin{center}
\includegraphics[scale=0.45]{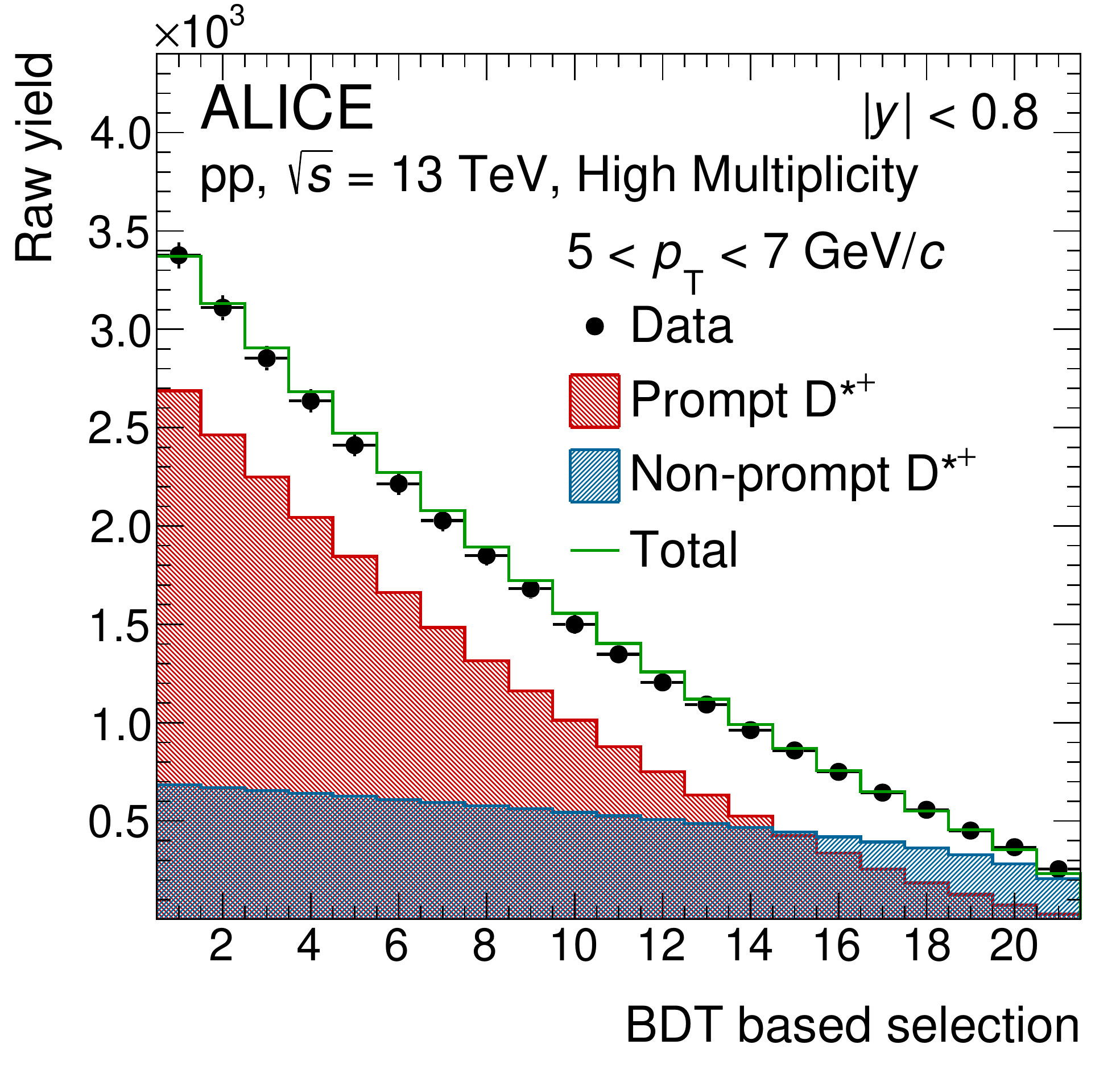}
\caption{Raw yield as a function of the BDT selection and extracted contributions from prompt and non-prompt \Dstar mesons with $5<\pt<7~\GeV/c$ and $|y|<0.8$ in the high-multiplicity triggered pp collisions at $\s=13~\TeV$.}
\label{fig:cutvar}
\end{center}
\end{figure}

The distribution was fitted with a combination of the Gaussian function corresponding to the \Dstar signal and a background function. The shape of the background distribution can be described with the function $p_0\sqrt{\Delta M - M_{\uppi}}\text{ }e^{p_1(\Delta M - M_{\uppi})}$, where $p_0$ and $p_1$ are free parameters and $M_{\uppi}$ is the pion mass. The corresponding \AccEff factors were obtained with MC simulations analogous to those used for the BDT trainings. Finally, the fraction of (non-)prompt \Dstar mesons for a given set of selections was computed as
\begin{equation}
    f_{\mathrm{(non\text{-})prompt},j} = \frac{(\AccEff)_{\mathrm{(non\text{-})prompt},j} \times N_\mathrm{(non\text{-})prompt}}{\effP{j}\times\Np+\effNP{j}\times\Nnp}.
\end{equation}
A different index "$j$" is used here to underline the fact that not all of these sets are used for the $\chi^2$ minimisation given by Eq.~\ref{eq:eq_set}. The uncertainty of $f_{\mathrm{(non\text{-})prompt},j}$ is computed considering the covariance matrix of \Np and \Nnp obtained in the $\chi^2$-minimisation procedure.
Figure~\ref{fig:cutvar} shows an example of a raw-yield extracted as a function of the BDT selection employed in the minimisation procedure for \Dstar mesons with $5<\pt<7~\GeV/c$ in high-multiplicity triggered pp collisions at $\s=13~\TeV$. The leftmost data point of the distribution is the raw yield corresponding to the loosest selection on the BDT output related to the candidate probability of being a non-prompt \Dstar meson, while the rightmost one corresponds to the strictest selection, which is expected to preferentially select non-prompt \Dstar mesons. The \AccEff factors obtained for these selections ranged from about 10\% to 0.1\% for prompt \Dstar mesons, and from about 30\% to 10\% for non-prompt \Dstar mesons. The prompt and non-prompt components are represented by the red and blue filled histograms, respectively, while their sum is depicted by the green histogram. The \fprompt factor obtained in the charm-enhanced sample is higher than 90\% for both the high-multiplicity and minimum-bias triggered events, while the \fnonprompt in the beauty-enhanced sample is $\sim 60\%$ and $\sim 40\%$ almost independent of \pt for high-multiplicity and minimum-bias triggered events, respectively. 

\begin{figure}[!t]
\begin{center}
\includegraphics[scale=0.8]{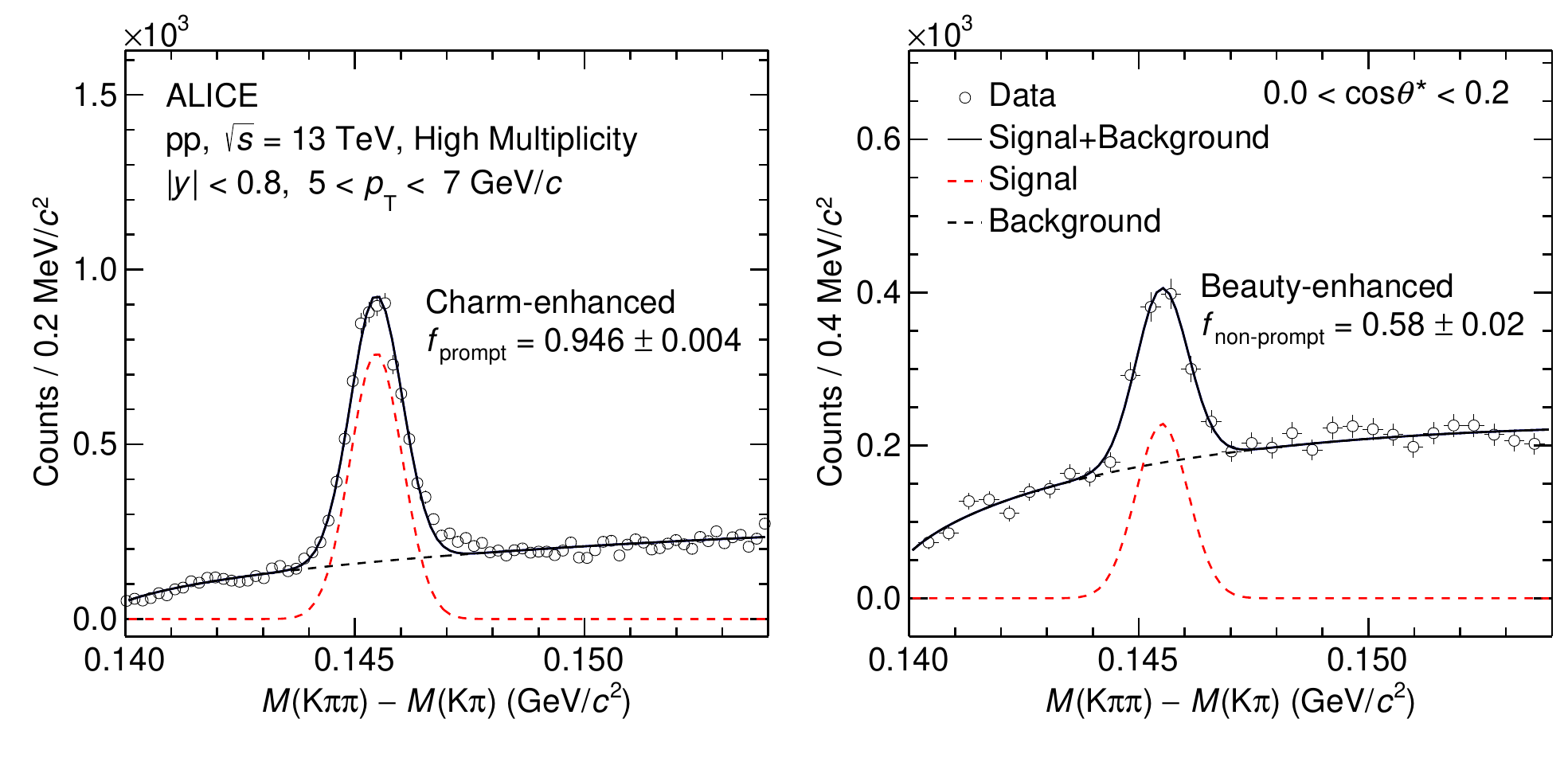}
\caption{Invariant-mass distributions $\Delta M$ for the charm-enhanced (left panel) and the beauty-enhanced (right panel) samples in pp collisions at $\s=13~\TeV$, selected with the high-multiplicity trigger. The distributions are obtained at midrapidity ($|y|<0.8$) in $5<\pt<7~\GeV/c$ and $0.0<\cost<0.2$.}
\label{fig:invmass}
\end{center}
\end{figure}

The raw yields of \Dstar mesons were extracted in different \cost and \pt intervals for both the beauty- and the charm-enhanced samples. Figure~\ref{fig:invmass} shows the $\Delta M$ distributions in $5<\pt<7~\GeV/c$ and $0.0<\cost<0.2$ intervals for the beauty- and charm-enhanced samples in high-multiplicity triggered pp collisions at $\s=13~\TeV$. The width of the Gaussian function for the \Dstar signal was fixed to the value from the MC simulation.

\begin{figure}[!b]
\begin{center}
\includegraphics[scale=0.8]{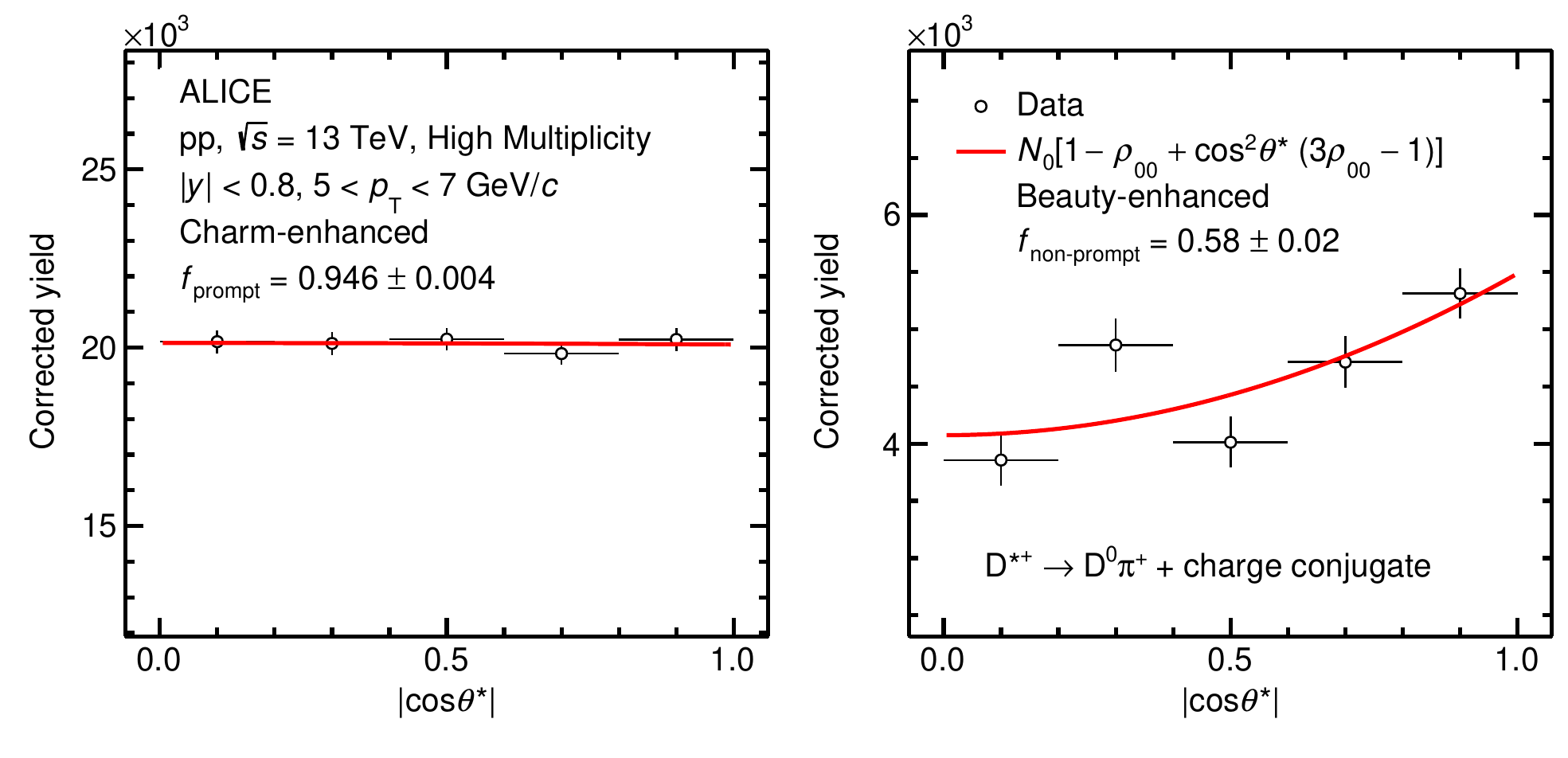}
\caption{Acceptance and efficiency corrected angular distributions of the decay product in the rest frame of the \Dstar meson with respect to the helicity axis. The left panel shows the angular distribution for the charm-enhanced \Dstar sample in pp collisions at $\s=13~\TeV$, selected with the high-multiplicity trigger and the right panel shows the same but for the beauty-enhanced \Dstar sample. The distributions are obtained at midrapidity ($|y|<0.8$) in $5<\pt<7~\GeV/c$ and are fitted with Eq.~\ref{eqn1} to extract the \rzz values. Only statistical uncertainties are shown.}
\label{costhetastardist}
\end{center}
\end{figure}

The raw yields were corrected for the corresponding \AccEff factors in each \cost and \pt intervals. The impact on the \rzz measurement due to the shape of the \cost and \pt distributions in the MC simulations was studied by weighting the generated MC distributions in order to reproduce the measured \cost and \pt distributions, and the effect on the \AccEff values was found to be smaller than 0.1\% in the analysed \pt intervals. The efficiency and acceptance corrected angular distributions at midrapidity for the selected \pt interval in high-multiplicity triggered pp collisions at $\s=13~\TeV$ are shown in Fig.~\ref{costhetastardist}. The left panel of Fig.~\ref{costhetastardist} shows the angular distribution of the charm-enhanced \Dstar sample and the right panel shows the same but for the beauty-enhanced \Dstar sample. The absolute value of \cost was considered due to the limited size of the analysed sample. The angular distributions were further fitted with the functional form given in Eq.~\ref{eqn1} to extract the \rzz values for both the charm- and the beauty-enhanced samples in each \pt interval.

\begin{figure}[!t]
\begin{center}
\includegraphics[scale=0.45]{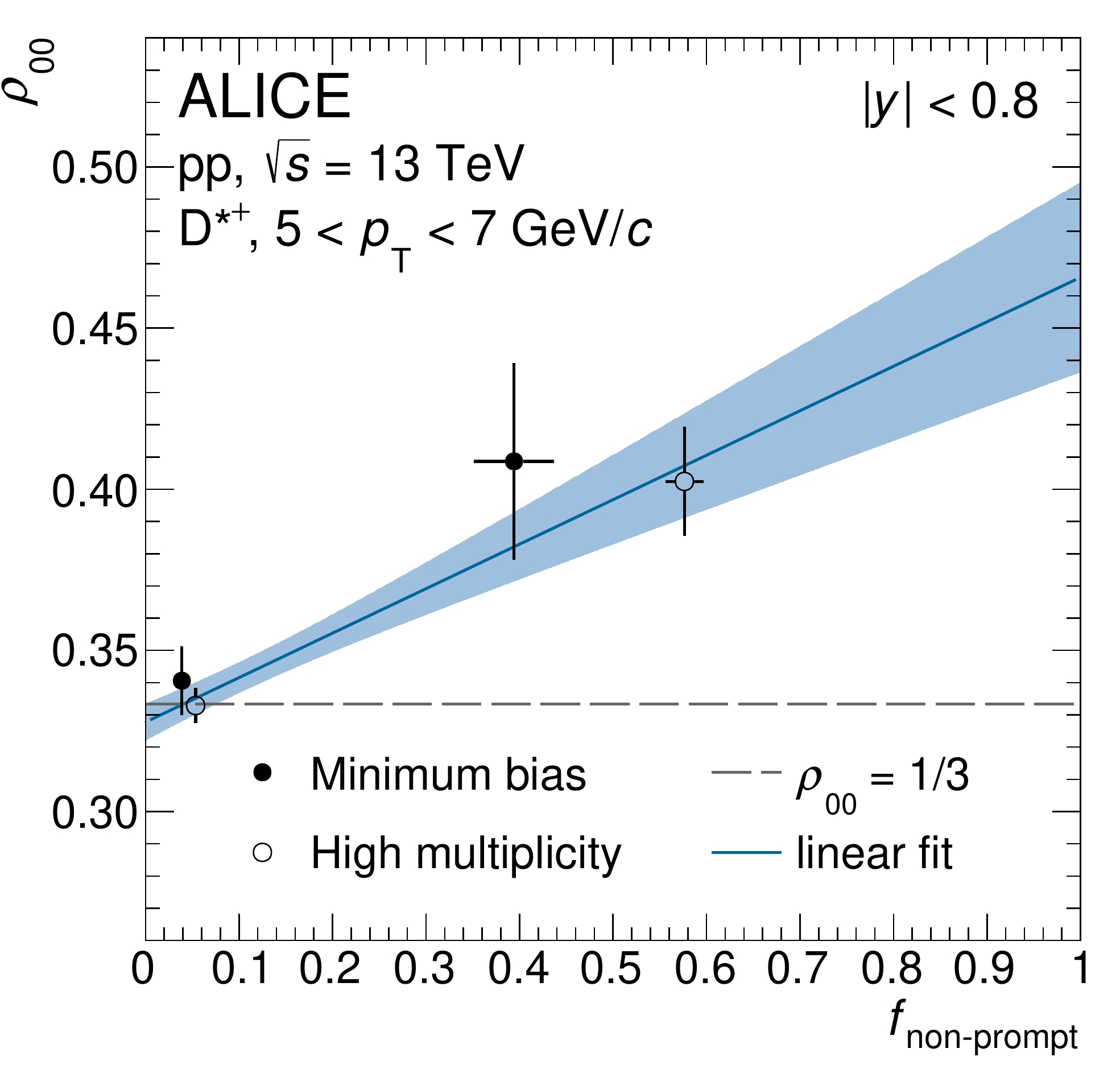}
\caption{The spin density matrix element (\rzz) as a function of \fnonprompt in $5<\pt<7~\GeV/c$ for charm- and beauty-enhanced samples obtained for minimum-bias and high-multiplicity triggered pp collisions at $\s=13~\TeV$. Only statistical uncertainties are shown. The blue line represents the linear fit to the data with the blue band indicating its $1\sigma$ confidence interval.} 
\label{fig:Rho00Extrap} 
\end{center}
\end{figure}

It was verified with a MC simulation that the \rzz parameter linearly depends on the fraction of non-prompt \Dstar meson in the analysed sample. Hence, a linear fit of the measured \rzz parameters obtained for the charm- and beauty-enhanced samples as a function of the \fnonprompt was performed to obtain the \rzz values for prompt and non-prompt \Dstar mesons. The charm- and the beauty-enhanced selections used in this analysis provided two samples of independent candidates. Moreover, the samples of candidates in the minimum-bias and high-multiplicity triggered events are statistically independent. Considering that the measured \rzz parameter can be decomposed as a linear combination of the prompt and non-prompt components, 
\begin{equation}
    \rzz = \fprompt \times \rzz^\mathrm{prompt} + \fnonprompt \times \rzz^\mathrm{non\text{-}prompt},
\end{equation}
the four independent measurements corresponding to the charm- and beauty-enhanced samples for both minimum-bias and high-multiplicity events were fitted with a linear function in order to extrapolate it to $\fnonprompt=0$ and $\fnonprompt=1$. A similar procedure was adopted in Ref.~\cite{CMS:2020qul} to measure the elliptic flow of prompt and non-prompt D mesons. With the current data samples it is not possible to perform the linear extrapolation of the \rzz parameter independently for the minimum-bias and high-multiplicity collisions, which may have different physical origin, and the reported results are for the combined sample. Figure~\ref{fig:Rho00Extrap} shows the linear fit of \rzz as a function of \fnonprompt in $5<\pt<7~\GeV/c$. The filled (empty) markers correspond to the measurements performed in the minimum-bias (high-multiplicity) pp collisions. The coloured band represents the $1\sigma$ confidence interval obtained from the linear fit. The \rzz parameter for prompt and non-prompt \Dstar mesons is then obtained by evaluating the fit function for $\fnonprompt=0$ and $\fnonprompt=1$, respectively. 

\section{Systematic uncertainties}
There are four major sources of systematic uncertainties in the measurement of \rzz for prompt and non-prompt \Dstar mesons. These uncertainties are related to: i) the signal extraction, ii) the track reconstruction and selection efficiencies, iii) the BDT selection efficiency, and iv) the (non-)prompt fraction estimation. The systematic uncertainty arising from the signal extraction was estimated by varying the invariant-mass fit range, leaving the Gaussian width free in the fit, and using a bin-counting method for the raw-yield extraction. In the bin-counting method the signal yield is extracted by integrating the invariant-mass distribution after subtracting the background estimated from the background fit function.
The systematic uncertainty associated with the track selection and reconstruction efficiency was evaluated by varying the track-quality selection criteria as described in Ref.~\cite{ALICE:2017olh}. The systematic uncertainty on the selection efficiency originates from imperfections in the description of the decay kinematics as well as the detector resolution and alignment in the simulation. It was estimated by comparing the \rzz parameters obtained for prompt and non-prompt \Dstar mesons with different BDT selection criteria. The uncertainty associated with the (non-)prompt fraction estimation was evaluated by varying the selection criteria described in Sec.~\ref{sec:analysis}, as well as the configuration of the invariant-mass fits.
All of the above mentioned variations were performed independently and, assuming they are uncorrelated with each other, added in quadrature to obtain the total systematic uncertainties. Table~\ref{tab:sysunc_yieldtable} summarises the values of the systematic uncertainties for the lowest and highest \pt intervals, reporting separately the various sources for prompt and non-prompt \Dstar mesons. The total systematic uncertainty ranges between 2.5\% and 3.5\% for prompt \Dstar mesons, and between 6.5\% and 8\% for non-prompt \Dstar mesons. These systematic uncertainties were considered as fully correlated as a function of \pt.

\begin{table}[!b]
\caption{Summary of the relative systematic uncertainties on the \rzz parameter of prompt and non-prompt \Dstar mesons for different \pt intervals.}
\centering
\renewcommand*{\arraystretch}{1.2}
\begin{tabular}[t]{l|>{\centering}p{0.10\linewidth}>{\centering}p{0.10\linewidth}|>{\centering}p{0.10\linewidth}>{\centering\arraybackslash}p{0.10\linewidth}}
\toprule
 & \multicolumn{2}{c|}{Prompt} & \multicolumn{2}{c}{Non-prompt} \\
\midrule
$\pt~(\GeV/c)$ & $5\mbox{--}7$ & $10\mbox{--}20$ & $5\mbox{--}7$ & $10\mbox{--}20$ \\
\midrule
Signal yield & 2.0\% & 0.5\% & 4.0\% & 2.0\% \\
Tracking efficiency & negl. & 1.0\% & 2.0\% & 4.0\% \\
BDT efficiency & 0.5\% & 3.5\% & 6.0\% & 5.0\% \\
(Non-)prompt fraction & negl. & negl. & 0.5\% & negl. \\
\midrule
Total & 2.5\% & 3.5\% & 8.0\% & 6.5\% \\
\bottomrule
\end{tabular}
\label{tab:sysunc_yieldtable}	
\end{table}

In order to validate the results of the measurements, an additional analysis was performed with respect to a random quantisation axis (random unit vector direction in 3-dimensional space) which breaks the correlation between the helicity axis and the daughter momentum direction, due to helicity conservation. As expected, the extracted \rzz values for both prompt and non-prompt \Dstar mesons are consistent with $1/3$. 

\section{Results}

\begin{figure}[!t]
\begin{center}
\includegraphics[scale=0.45]{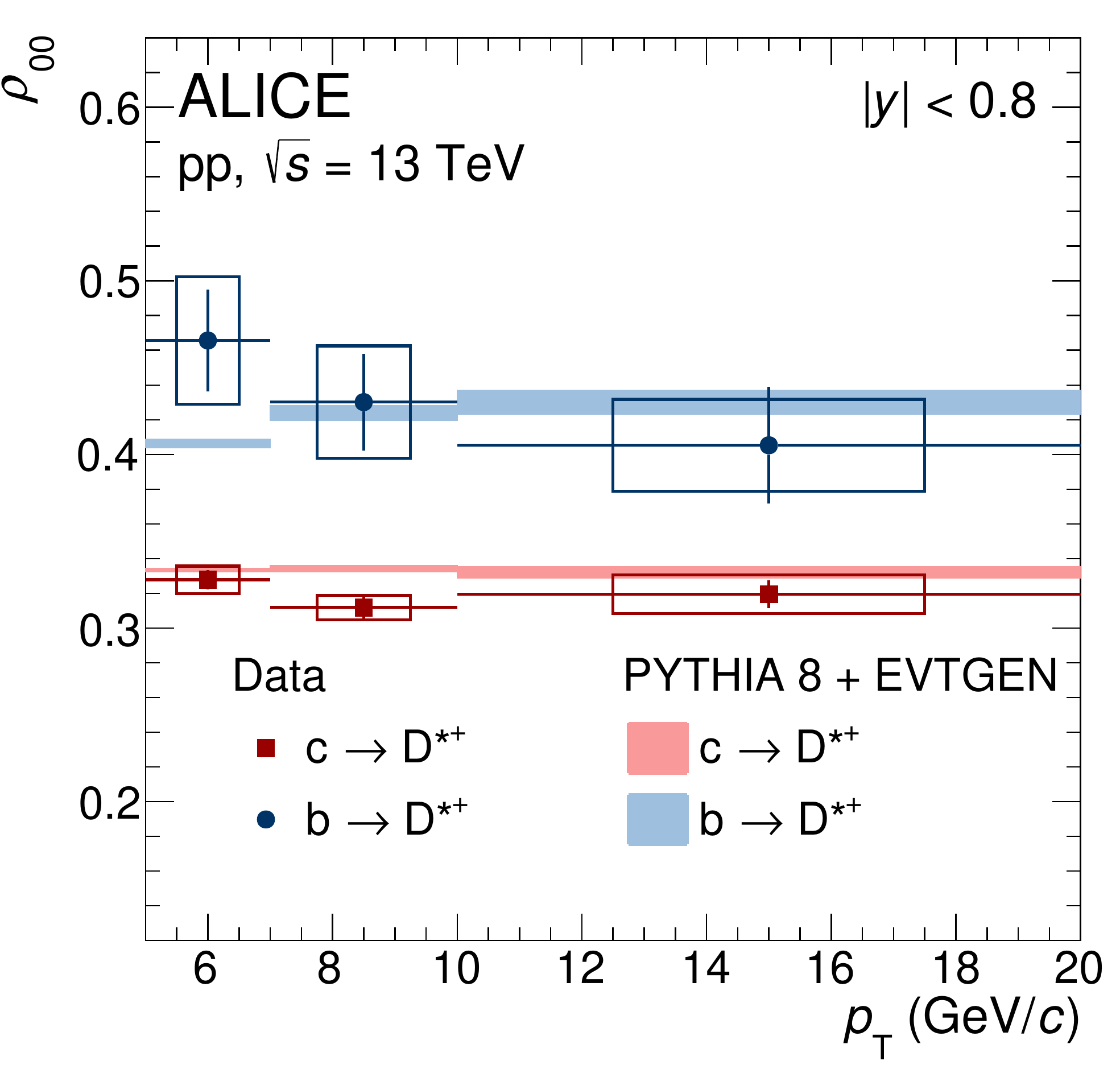}
\caption{The spin density matrix element (\rzz) for prompt and non-prompt \Dstar mesons as a function of \pt in pp collisions at $\s=13~\TeV$. Measurements are carried out with respect to the helicity axis at $|y|<0.8$. Statistical and systematic uncertainties are represented by bars and boxes, respectively. The measurements are compared with the estimations from the QCD-based MC event generator \pythia + \evtgen~\cite{Skands:2014pea,Ryd:2005zz}. Model estimations are shown by colour bands where the width of the band corresponds to the statistical uncertainty in model calculations.} 
\label{result} 
\end{center}
\end{figure}

Figure~\ref{result} shows the \rzz values with respect to the helicity axis for prompt and non-prompt \Dstar mesons as a function of \pt in pp collisions at $\s=13~\TeV$. The measurements are performed in $|y|<0.8$. The \rzz values for prompt \Dstar mesons are consistent with $1/3$, while for non-prompt \Dstar mesons they are significantly larger than $1/3$ throughout the studied \pt range. The values measured in the three $\pt$ intervals were averaged using the corrected yields of prompt and non-prompt \Dstar mesons as weights. The resulting \rzz values in the \pt interval between $5$ and $20~\GeV/c$ are
\begin{equation}
\begin{aligned}
\rzz(\text{prompt }\Dstar) &= 0.324\pm0.004(\text{stat.})\pm0.008(\text{syst.}),\\
    \rzz(\text{non-prompt }\Dstar) &= 0.455\pm0.022(\text{stat.})\pm0.035(\text{syst.}).
    \end{aligned}
\end{equation}

This finding implies no spin alignment for prompt \Dstar mesons with a simultaneous non-zero spin alignment of non-prompt \Dstar mesons in high-energy pp collisions. The evidence of non-prompt \Dstar-meson spin alignment can be understood as a consequence of helicity conservation in the decay of beauty scalar mesons to vector mesons. Measured \rzz values are further compared with the MC event generator \pythia + \evtgen~\cite{Skands:2014pea,Ryd:2005zz}. The \evtgen decay package, which conserves the helicity and account for the V$-$A nature in the decay of beauty mesons, is used in place of the default \pythia decayer. In this case, model calculations are found to be in agreement with the extracted \rzz values for both prompt and non-prompt \Dstar mesons, while if the \evtgen package is not used, the \rzz parameter is found to be compatible with $1/3$ for both prompt and non-prompt \Dstar mesons. The measured \rzz for non-prompt \Dstar mesons is qualitatively consistent with the longitudinal polarisation fraction ($f_\mathrm{L}$) observed by the LHCb collaboration in the $\mathrm{B^0}\rightarrow \mathrm{D^{*-}}\mathrm{D_{s}^{*+}}$ decay with respect to a different quantisation axis defined by the B-meson momentum direction~\cite{LHCb:2021sqa}. The direct comparison between the magnitudes of $f_\mathrm{L}$ and \rzz measured by the LHCb and ALICE Collaborations is not possible due to the different quantisation axis and the different decay channels. Therefore, dedicated model calculations are required for a quantitative comparison of the ALICE and LHCb data.

\begin{figure}[!t]
\begin{center}
\includegraphics[scale=0.5]{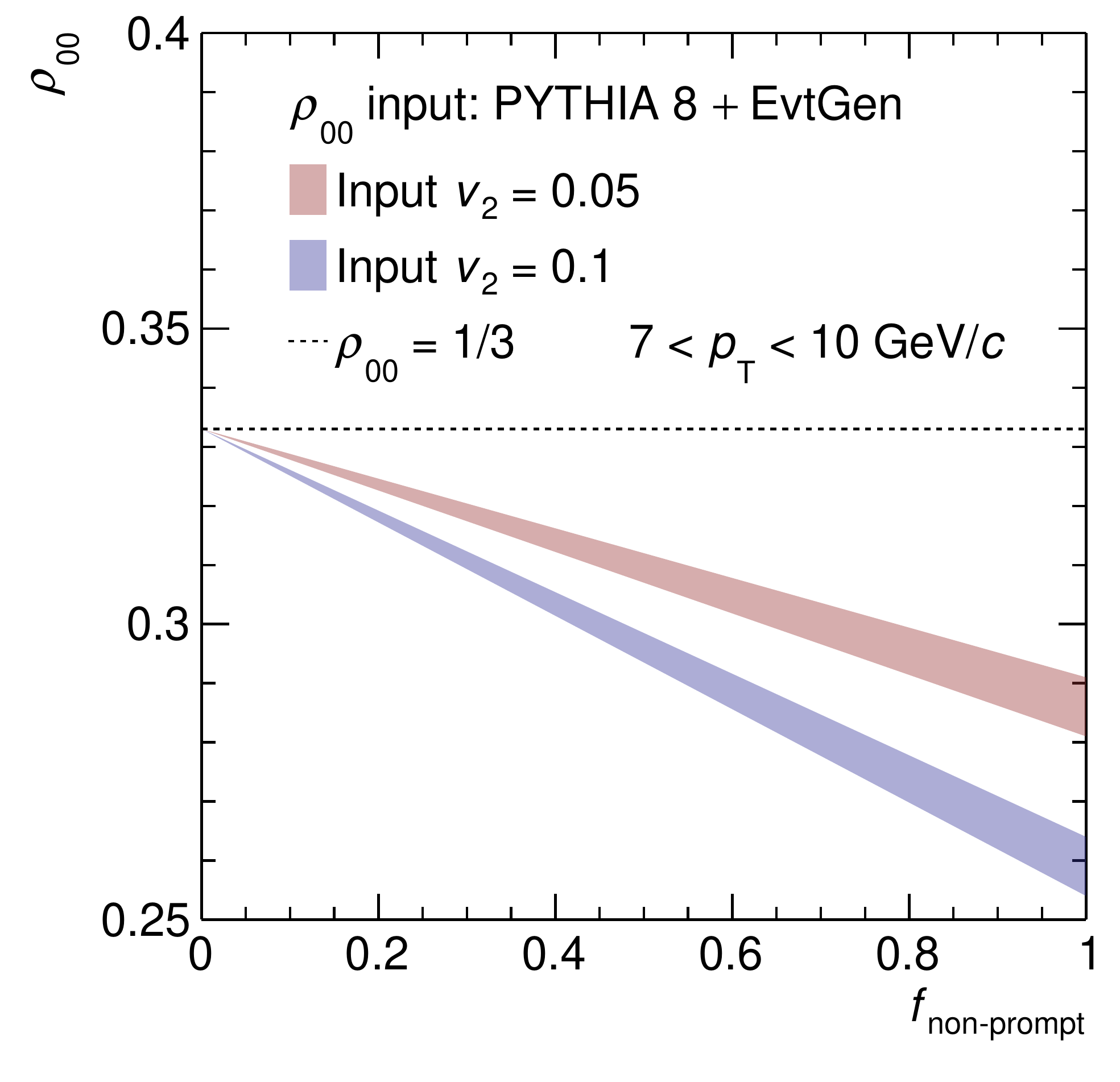}
\caption{Model simulation based on the MC event generator \pythia + \evtgen~\cite{Skands:2014pea,Ryd:2005zz} for the spin density matrix element (\rzz) for \Dstar mesons with respect to the direction perpendicular to the reaction plane as a function of \fnonprompt. Model simulations are carried out for \Dstar mesons at midrapidity ($|y|<0.8$) with two input values of $v_{2}$.} 
\label{result2} 
\end{center}
\end{figure}

After the validation of the helicity conservation in the \HbtoDstar decays via the comparison with the measurement presented in this letter, the predicted \rzz value obtained with \pythia and the \evtgen decay package can be used to estimate the magnitude of the corresponding contribution to the spin alignment in heavy-ion collisions. In particular, the spin alignment of non-prompt \Dstar meson with respect to the helicity axis can affect the measurements of \Dstar spin alignment with respect to the reaction plane in heavy-ion collisions. Due to the elliptic flow of the non-prompt \Dstar mesons~\cite{ALICE:2017pbx,CMS:2017vhp,ALICE:2018gif,ALICE:2020iug,CMS:2020bnz} in heavy-ion collisions, the helicity axis is correlated to the direction of the angular momentum and the magnetic field, which are perpendicular to the orientation of the reaction plane~\cite{ALICE:2020iug}. In order to obtain a quantitative estimate of this effect for heavy-ion collisions, a combination of \pythia and the \evtgen decay package, which are in good agreement with ALICE measurements as shown in Fig.~\ref{result}, is used to model the spin alignment of non-prompt \Dstar-mesons with respect to the helicity axis.
Elliptic flow modulations of non-prompt \Dstar mesons are introduced in the simulation by an appropriate rotation of the \Dstar and its decay product momentum in the azimuthal plane. Introducing $v_{2}$ modulation aligns the momentum direction of non-prompt \Dstar in the direction perpendicular to the reaction plane.
Two different calculations with $v_{2}$ values of 0.05 and 0.1 were performed. These values were chosen to mimic measurements reported by the ALICE and ATLAS Collaborations of leptons from beauty-hadron decays for midcentral Pb--Pb collisions~\cite{ALICE:2020hdw, ATLAS:2020yxw}. Figure~\ref{result2} shows the calculated \rzz values for \Dstar mesons as a function of \fnonprompt with respect to the direction perpendicular to the reaction plane. It is found that in the presence of $v_{2}$, helicity conservation in the beauty-meson decay leads to a value of \rzz less than $1/3$. The opposite direction with respect to $1/3$ compared to the measurement presented in this letter is due to the different quantisation axis.
In the recombination hadronisation scenario of polarised quarks, the expected value of \rzz for both the magnetic field and angular momentum is also less than $1/3$~\cite{Yang:2017sdk}. Therefore, in the presence of $v_{2}$ the spin alignment of non-prompt \Dstar mesons due to the helicity conservation in beauty-meson decays mimics the signal of global spin alignment, expected due to the presence of large angular momentum and magnetic field in midcentral heavy-ion collisions. In the presence of additional non-flow correlations (i.e.~correlations not induced by the collective expansion but rather by decays and jet contributions)~\cite{Borghini:2000cm}, this effect can get further enhanced. A sizeable fraction of non-prompt \Dstar mesons, e.g.~$\fnonprompt = 0.2$~\cite{ALICE:2021mgk,ALICE:2022tji} can give on the order of 5\% deviation from the $\rzz = 1/3$. This contribution is not negligible considering the recent \Jpsi polarisation measurement in Pb--Pb collisions by the ALICE Collaboration where a deviation up to 10\% from the $\rzz = 1/3$ was observed~\cite{ALICE:2022sli}. Thanks to the higher luminosity and improved spatial resolution of the upgraded ITS installed in ALICE for the LHC Run 3~\cite{ALICE:2013nwm,ALICE:2023udb}, future measurements with non-prompt charm hadrons of the spin alignment in pp collisions and azimuthal anisotropy in heavy-ion collisions will provide a more accurate baseline for understanding effects induced by the initial angular momentum and magnetic field in non-central heavy-ion collisions.

\section{Summary}
The first measurements for prompt and non-prompt \Dstar meson spin alignment with respect to the helicity axis at midrapidity in pp collisions at $\s = 13~\TeV$ are presented. No spin alignment is observed for the prompt \Dstar mesons in pp collisions, with measured \rzz values being consistent with $1/3$. This observation is consistent with previous measurements in \ee collisions far from the \ZZero resonance and suggests that charm quarks are either produced unpolarised or their polarisation is washed out during the hadronisation process. The measured value for the non-prompt \Dstar-meson spin density element is $\rzz=0.455\pm0.022(\text{stat.})\pm0.035(\text{syst.})$ in the $5<\pt<20~\GeV/c$ interval. The measured value is compatible with the prediction from \pythia + \evtgen simulations and is interpreted as an evidence of spin alignment of non-prompt \Dstar mesons with respect to the helicity axis due to the helicity conservation in beauty-meson decays. Based on the measurements and MC simulations, it was estimated that the spin alignment of the non-prompt \Dstar mesons in the presence of
elliptic flow can mimic the signal of global spin alignment in heavy-ion collisions. The new data presented in this letter provide an important baseline for future spin alignment measurements of \Dstar vector mesons in heavy-ion collisions, which probe the initial angular momentum and magnetic field.

\newenvironment{acknowledgement}{\relax}{\relax}
\begin{acknowledgement}
\section*{Acknowledgements}
% Version: 2022-11-28

The ALICE Collaboration would like to thank all its engineers and technicians for their invaluable contributions to the construction of the experiment and the CERN accelerator teams for the outstanding performance of the LHC complex.
The ALICE Collaboration gratefully acknowledges the resources and support provided by all Grid centres and the Worldwide LHC Computing Grid (WLCG) collaboration.
The ALICE Collaboration acknowledges the following funding agencies for their support in building and running the ALICE detector:
A. I. Alikhanyan National Science Laboratory (Yerevan Physics Institute) Foundation (ANSL), State Committee of Science and World Federation of Scientists (WFS), Armenia;
Austrian Academy of Sciences, Austrian Science Fund (FWF): [M 2467-N36] and Nationalstiftung f\"{u}r Forschung, Technologie und Entwicklung, Austria;
Ministry of Communications and High Technologies, National Nuclear Research Center, Azerbaijan;
Conselho Nacional de Desenvolvimento Cient\'{\i}fico e Tecnol\'{o}gico (CNPq), Financiadora de Estudos e Projetos (Finep), Funda\c{c}\~{a}o de Amparo \`{a} Pesquisa do Estado de S\~{a}o Paulo (FAPESP) and Universidade Federal do Rio Grande do Sul (UFRGS), Brazil;
Bulgarian Ministry of Education and Science, within the National Roadmap for Research Infrastructures 2020-2027 (object CERN), Bulgaria;
Ministry of Education of China (MOEC) , Ministry of Science \& Technology of China (MSTC) and National Natural Science Foundation of China (NSFC), China;
Ministry of Science and Education and Croatian Science Foundation, Croatia;
Centro de Aplicaciones Tecnol\'{o}gicas y Desarrollo Nuclear (CEADEN), Cubaenerg\'{\i}a, Cuba;
Ministry of Education, Youth and Sports of the Czech Republic, Czech Republic;
The Danish Council for Independent Research | Natural Sciences, the VILLUM FONDEN and Danish National Research Foundation (DNRF), Denmark;
Helsinki Institute of Physics (HIP), Finland;
Commissariat \`{a} l'Energie Atomique (CEA) and Institut National de Physique Nucl\'{e}aire et de Physique des Particules (IN2P3) and Centre National de la Recherche Scientifique (CNRS), France;
Bundesministerium f\"{u}r Bildung und Forschung (BMBF) and GSI Helmholtzzentrum f\"{u}r Schwerionenforschung GmbH, Germany;
General Secretariat for Research and Technology, Ministry of Education, Research and Religions, Greece;
National Research, Development and Innovation Office, Hungary;
Department of Atomic Energy Government of India (DAE), Department of Science and Technology, Government of India (DST), University Grants Commission, Government of India (UGC) and Council of Scientific and Industrial Research (CSIR), India;
National Research and Innovation Agency - BRIN, Indonesia;
Istituto Nazionale di Fisica Nucleare (INFN), Italy;
Japanese Ministry of Education, Culture, Sports, Science and Technology (MEXT) and Japan Society for the Promotion of Science (JSPS) KAKENHI, Japan;
Consejo Nacional de Ciencia (CONACYT) y Tecnolog\'{i}a, through Fondo de Cooperaci\'{o}n Internacional en Ciencia y Tecnolog\'{i}a (FONCICYT) and Direcci\'{o}n General de Asuntos del Personal Academico (DGAPA), Mexico;
Nederlandse Organisatie voor Wetenschappelijk Onderzoek (NWO), Netherlands;
The Research Council of Norway, Norway;
Commission on Science and Technology for Sustainable Development in the South (COMSATS), Pakistan;
Pontificia Universidad Cat\'{o}lica del Per\'{u}, Peru;
Ministry of Education and Science, National Science Centre and WUT ID-UB, Poland;
Korea Institute of Science and Technology Information and National Research Foundation of Korea (NRF), Republic of Korea;
Ministry of Education and Scientific Research, Institute of Atomic Physics, Ministry of Research and Innovation and Institute of Atomic Physics and University Politehnica of Bucharest, Romania;
Ministry of Education, Science, Research and Sport of the Slovak Republic, Slovakia;
National Research Foundation of South Africa, South Africa;
Swedish Research Council (VR) and Knut \& Alice Wallenberg Foundation (KAW), Sweden;
European Organization for Nuclear Research, Switzerland;
Suranaree University of Technology (SUT), National Science and Technology Development Agency (NSTDA) and National Science, Research and Innovation Fund (NSRF via PMU-B B05F650021), Thailand;
Turkish Energy, Nuclear and Mineral Research Agency (TENMAK), Turkey;
National Academy of  Sciences of Ukraine, Ukraine;
Science and Technology Facilities Council (STFC), United Kingdom;
National Science Foundation of the United States of America (NSF) and United States Department of Energy, Office of Nuclear Physics (DOE NP), United States of America.
In addition, individual groups or members have received support from:
European Research Council, Strong 2020 - Horizon 2020, Marie Sk\l{}odowska Curie (grant nos. 950692, 824093, 896850), European Union;
Academy of Finland (Center of Excellence in Quark Matter) (grant nos. 346327, 346328), Finland;
Programa de Apoyos para la Superaci\'{o}n del Personal Acad\'{e}mico, UNAM, Mexico.
%%%%%%% done by webmaster team
\end{acknowledgement}
\bibliographystyle{utphys}
%\bibliography{biblio}
\bibliography{Dstar_Spin_alignment}
    
\newpage
\appendix
\section{The ALICE Collaboration}
\label{app:collab}
% ALICE Collaboration author list for 2022-11-28
\begin{flushleft} 
\small

S.~Acharya\,\orcidlink{0000-0002-9213-5329}\,$^{\rm 124}$, 
D.~Adamov\'{a}\,\orcidlink{0000-0002-0504-7428}\,$^{\rm 85}$, 
A.~Adler$^{\rm 69}$, 
G.~Aglieri Rinella\,\orcidlink{0000-0002-9611-3696}\,$^{\rm 32}$, 
M.~Agnello\,\orcidlink{0000-0002-0760-5075}\,$^{\rm 29}$, 
N.~Agrawal\,\orcidlink{0000-0003-0348-9836}\,$^{\rm 50}$, 
Z.~Ahammed\,\orcidlink{0000-0001-5241-7412}\,$^{\rm 132}$, 
S.~Ahmad\,\orcidlink{0000-0003-0497-5705}\,$^{\rm 15}$, 
S.U.~Ahn\,\orcidlink{0000-0001-8847-489X}\,$^{\rm 70}$, 
I.~Ahuja\,\orcidlink{0000-0002-4417-1392}\,$^{\rm 37}$
% ...

\section*{Affiliation Notes}

$^{\rm I}$ Deceased\\
$^{\rm II}$ Also at: Max-Planck-Institut f\"{u}r Physik, Munich, Germany\\
$^{\rm III}$ Also at: Italian National Agency for New Technologies, Energy and Sustainable Economic Development (ENEA), Bologna, Italy\\
$^{\rm IV}$ Also at: Dipartimento DET del Politecnico di Torino, Turin, Italy\\
$^{\rm V}$ Also at: Department of Applied Physics, Aligarh Muslim University, Aligarh, India\\
$^{\rm VI}$ Also at: Institute of Theoretical Physics, University of Wroclaw, Poland\\
$^{\rm VII}$ Also at: An institution covered by a cooperation agreement with CERN\\

\section*{Collaboration Institutes}

$^{1}$ A.I. Alikhanyan National Science Laboratory (Yerevan Physics Institute) Foundation, Yerevan, Armenia\\
$^{2}$ AGH University of Krakow, Cracow, Poland\\
$^{3}$ Bogolyubov Institute for Theoretical Physics, National Academy of Sciences of Ukraine, Kiev, Ukraine\\
$^{4}$ Bose Institute, Department of Physics  and Centre for Astroparticle Physics and Space Science (CAPSS), Kolkata, India\\
$^{5}$ California Polytechnic State University, San Luis Obispo, California, United States\\
$^{6}$ Central China Normal University, Wuhan, China\\
$^{7}$ Centro de Aplicaciones Tecnol\'{o}gicas y Desarrollo Nuclear (CEADEN), Havana, Cuba\\
$^{8}$ Centro de Investigaci\'{o}n y de Estudios Avanzados (CINVESTAV), Mexico City and M\'{e}rida, Mexico\\
$^{9}$ Chicago State University, Chicago, Illinois, United States\\
$^{10}$ China Institute of Atomic Energy, Beijing, China\\
$^{11}$ Chungbuk National University, Cheongju, Republic of Korea\\
$^{12}$ Comenius University Bratislava, Faculty of Mathematics, Physics and Informatics, Bratislava, Slovak Republic\\
$^{13}$ COMSATS University Islamabad, Islamabad, Pakistan\\
$^{14}$ Creighton University, Omaha, Nebraska, United States\\
$^{15}$ Department of Physics, Aligarh Muslim University, Aligarh, India\\
$^{16}$ Department of Physics, Pusan National University, Pusan, Republic of Korea\\
$^{17}$ Department of Physics, Sejong University, Seoul, Republic of Korea\\
$^{18}$ Department of Physics, University of California, Berkeley, California, United States\\
$^{19}$ Department of Physics, University of Oslo, Oslo, Norway\\
$^{20}$ Department of Physics and Technology, University of Bergen, Bergen, Norway\\
$^{21}$ Dipartimento di Fisica, Universit\`{a} di Pavia, Pavia, Italy\\
$^{22}$ Dipartimento di Fisica dell'Universit\`{a} and Sezione INFN, Cagliari, Italy\\
$^{23}$ Dipartimento di Fisica dell'Universit\`{a} and Sezione INFN, Trieste, Italy\\
$^{24}$ Dipartimento di Fisica dell'Universit\`{a} and Sezione INFN, Turin, Italy\\
$^{25}$ Dipartimento di Fisica e Astronomia dell'Universit\`{a} and Sezione INFN, Bologna, Italy\\
$^{26}$ Dipartimento di Fisica e Astronomia dell'Universit\`{a} and Sezione INFN, Catania, Italy\\
$^{27}$ Dipartimento di Fisica e Astronomia dell'Universit\`{a} and Sezione INFN, Padova, Italy\\
$^{28}$ Dipartimento di Fisica `E.R.~Caianiello' dell'Universit\`{a} and Gruppo Collegato INFN, Salerno, Italy\\
$^{29}$ Dipartimento DISAT del Politecnico and Sezione INFN, Turin, Italy\\
$^{30}$ Dipartimento di Scienze MIFT, Universit\`{a} di Messina, Messina, Italy\\
$^{31}$ Dipartimento Interateneo di Fisica `M.~Merlin' and Sezione INFN, Bari, Italy\\
$^{32}$ European Organization for Nuclear Research (CERN), Geneva, Switzerland\\
$^{33}$ Faculty of Electrical Engineering, Mechanical Engineering and Naval Architecture, University of Split, Split, Croatia\\
$^{34}$ Faculty of Engineering and Science, Western Norway University of Applied Sciences, Bergen, Norway\\
$^{35}$ Faculty of Nuclear Sciences and Physical Engineering, Czech Technical University in Prague, Prague, Czech Republic\\
$^{36}$ Faculty of Physics, Sofia University, Sofia, Bulgaria\\
$^{37}$ Faculty of Science, P.J.~\v{S}af\'{a}rik University, Ko\v{s}ice, Slovak Republic\\
$^{38}$ Frankfurt Institute for Advanced Studies, Johann Wolfgang Goethe-Universit\"{a}t Frankfurt, Frankfurt, Germany\\
$^{39}$ Fudan University, Shanghai, China\\
$^{40}$ Gangneung-Wonju National University, Gangneung, Republic of Korea\\
$^{41}$ Gauhati University, Department of Physics, Guwahati, India\\
$^{42}$ Helmholtz-Institut f\"{u}r Strahlen- und Kernphysik, Rheinische Friedrich-Wilhelms-Universit\"{a}t Bonn, Bonn, Germany\\
$^{43}$ Helsinki Institute of Physics (HIP), Helsinki, Finland\\
$^{44}$ High Energy Physics Group,  Universidad Aut\'{o}noma de Puebla, Puebla, Mexico\\
$^{45}$ Horia Hulubei National Institute of Physics and Nuclear Engineering, Bucharest, Romania\\
$^{46}$ Indian Institute of Technology Bombay (IIT), Mumbai, India\\
$^{47}$ Indian Institute of Technology Indore, Indore, India\\
$^{48}$ INFN, Laboratori Nazionali di Frascati, Frascati, Italy\\
$^{49}$ INFN, Sezione di Bari, Bari, Italy\\
$^{50}$ INFN, Sezione di Bologna, Bologna, Italy\\
$^{51}$ INFN, Sezione di Cagliari, Cagliari, Italy\\
$^{52}$ INFN, Sezione di Catania, Catania, Italy\\
$^{53}$ INFN, Sezione di Padova, Padova, Italy\\
$^{54}$ INFN, Sezione di Pavia, Pavia, Italy\\
$^{55}$ INFN, Sezione di Torino, Turin, Italy\\
$^{56}$ INFN, Sezione di Trieste, Trieste, Italy\\
$^{57}$ Inha University, Incheon, Republic of Korea\\
$^{58}$ Institute for Gravitational and Subatomic Physics (GRASP), Utrecht University/Nikhef, Utrecht, Netherlands\\
$^{59}$ Institute of Experimental Physics, Slovak Academy of Sciences, Ko\v{s}ice, Slovak Republic\\
$^{60}$ Institute of Physics, Homi Bhabha National Institute, Bhubaneswar, India\\
$^{61}$ Institute of Physics of the Czech Academy of Sciences, Prague, Czech Republic\\
$^{62}$ Institute of Space Science (ISS), Bucharest, Romania\\
$^{63}$ Institut f\"{u}r Kernphysik, Johann Wolfgang Goethe-Universit\"{a}t Frankfurt, Frankfurt, Germany\\
$^{64}$ Instituto de Ciencias Nucleares, Universidad Nacional Aut\'{o}noma de M\'{e}xico, Mexico City, Mexico\\
$^{65}$ Instituto de F\'{i}sica, Universidade Federal do Rio Grande do Sul (UFRGS), Porto Alegre, Brazil\\
$^{66}$ Instituto de F\'{\i}sica, Universidad Nacional Aut\'{o}noma de M\'{e}xico, Mexico City, Mexico\\
$^{67}$ iThemba LABS, National Research Foundation, Somerset West, South Africa\\
$^{68}$ Jeonbuk National University, Jeonju, Republic of Korea\\
$^{69}$ Johann-Wolfgang-Goethe Universit\"{a}t Frankfurt Institut f\"{u}r Informatik, Fachbereich Informatik und Mathematik, Frankfurt, Germany\\
$^{70}$ Korea Institute of Science and Technology Information, Daejeon, Republic of Korea\\
$^{71}$ KTO Karatay University, Konya, Turkey\\
$^{72}$ Laboratoire de Physique Subatomique et de Cosmologie, Universit\'{e} Grenoble-Alpes, CNRS-IN2P3, Grenoble, France\\
$^{73}$ Lawrence Berkeley National Laboratory, Berkeley, California, United States\\
$^{74}$ Lund University Department of Physics, Division of Particle Physics, Lund, Sweden\\
$^{75}$ Nagasaki Institute of Applied Science, Nagasaki, Japan\\
$^{76}$ Nara Women{'}s University (NWU), Nara, Japan\\
$^{77}$ National and Kapodistrian University of Athens, School of Science, Department of Physics , Athens, Greece\\
$^{78}$ National Centre for Nuclear Research, Warsaw, Poland\\
$^{79}$ National Institute of Science Education and Research, Homi Bhabha National Institute, Jatni, India\\
$^{80}$ National Nuclear Research Center, Baku, Azerbaijan\\
$^{81}$ National Research and Innovation Agency - BRIN, Jakarta, Indonesia\\
$^{82}$ Niels Bohr Institute, University of Copenhagen, Copenhagen, Denmark\\
$^{83}$ Nikhef, National institute for subatomic physics, Amsterdam, Netherlands\\
$^{84}$ Nuclear Physics Group, STFC Daresbury Laboratory, Daresbury, United Kingdom\\
$^{85}$ Nuclear Physics Institute of the Czech Academy of Sciences, Husinec-\v{R}e\v{z}, Czech Republic\\
$^{86}$ Oak Ridge National Laboratory, Oak Ridge, Tennessee, United States\\
$^{87}$ Ohio State University, Columbus, Ohio, United States\\
$^{88}$ Physics department, Faculty of science, University of Zagreb, Zagreb, Croatia\\
$^{89}$ Physics Department, Panjab University, Chandigarh, India\\
$^{90}$ Physics Department, University of Jammu, Jammu, India\\
$^{91}$ Physics Program and International Institute for Sustainability with Knotted Chiral Meta Matter (SKCM2), Hiroshima University, Hiroshima, Japan\\
$^{92}$ Physikalisches Institut, Eberhard-Karls-Universit\"{a}t T\"{u}bingen, T\"{u}bingen, Germany\\
$^{93}$ Physikalisches Institut, Ruprecht-Karls-Universit\"{a}t Heidelberg, Heidelberg, Germany\\
$^{94}$ Physik Department, Technische Universit\"{a}t M\"{u}nchen, Munich, Germany\\
$^{95}$ Politecnico di Bari and Sezione INFN, Bari, Italy\\
$^{96}$ Research Division and ExtreMe Matter Institute EMMI, GSI Helmholtzzentrum f\"ur Schwerionenforschung GmbH, Darmstadt, Germany\\
$^{97}$ Saga University, Saga, Japan\\
$^{98}$ Saha Institute of Nuclear Physics, Homi Bhabha National Institute, Kolkata, India\\
$^{99}$ School of Physics and Astronomy, University of Birmingham, Birmingham, United Kingdom\\
$^{100}$ Secci\'{o}n F\'{\i}sica, Departamento de Ciencias, Pontificia Universidad Cat\'{o}lica del Per\'{u}, Lima, Peru\\
$^{101}$ Stefan Meyer Institut f\"{u}r Subatomare Physik (SMI), Vienna, Austria\\
$^{102}$ SUBATECH, IMT Atlantique, Nantes Universit\'{e}, CNRS-IN2P3, Nantes, France\\
$^{103}$ Sungkyunkwan University, Suwon City, Republic of Korea\\
$^{104}$ Suranaree University of Technology, Nakhon Ratchasima, Thailand\\
$^{105}$ Technical University of Ko\v{s}ice, Ko\v{s}ice, Slovak Republic\\
$^{106}$ The Henryk Niewodniczanski Institute of Nuclear Physics, Polish Academy of Sciences, Cracow, Poland\\
$^{107}$ The University of Texas at Austin, Austin, Texas, United States\\
$^{108}$ Universidad Aut\'{o}noma de Sinaloa, Culiac\'{a}n, Mexico\\
$^{109}$ Universidade de S\~{a}o Paulo (USP), S\~{a}o Paulo, Brazil\\
$^{110}$ Universidade Estadual de Campinas (UNICAMP), Campinas, Brazil\\
$^{111}$ Universidade Federal do ABC, Santo Andre, Brazil\\
$^{112}$ University of Cape Town, Cape Town, South Africa\\
$^{113}$ University of Houston, Houston, Texas, United States\\
$^{114}$ University of Jyv\"{a}skyl\"{a}, Jyv\"{a}skyl\"{a}, Finland\\
$^{115}$ University of Kansas, Lawrence, Kansas, United States\\
$^{116}$ University of Liverpool, Liverpool, United Kingdom\\
$^{117}$ University of Science and Technology of China, Hefei, China\\
$^{118}$ University of South-Eastern Norway, Kongsberg, Norway\\
$^{119}$ University of Tennessee, Knoxville, Tennessee, United States\\
$^{120}$ University of the Witwatersrand, Johannesburg, South Africa\\
$^{121}$ University of Tokyo, Tokyo, Japan\\
$^{122}$ University of Tsukuba, Tsukuba, Japan\\
$^{123}$ University Politehnica of Bucharest, Bucharest, Romania\\
$^{124}$ Universit\'{e} Clermont Auvergne, CNRS/IN2P3, LPC, Clermont-Ferrand, France\\
$^{125}$ Universit\'{e} de Lyon, CNRS/IN2P3, Institut de Physique des 2 Infinis de Lyon, Lyon, France\\
$^{126}$ Universit\'{e} de Strasbourg, CNRS, IPHC UMR 7178, F-67000 Strasbourg, France, Strasbourg, France\\
$^{127}$ Universit\'{e} Paris-Saclay, Centre d'Etudes de Saclay (CEA), IRFU, D\'{e}partment de Physique Nucl\'{e}aire (DPhN), Saclay, France\\
$^{128}$ Universit\'{e}  Paris-Saclay, CNRS/IN2P3, IJCLab, Orsay, France\\
$^{129}$ Universit\`{a} degli Studi di Foggia, Foggia, Italy\\
$^{130}$ Universit\`{a} del Piemonte Orientale, Vercelli, Italy\\
$^{131}$ Universit\`{a} di Brescia, Brescia, Italy\\
$^{132}$ Variable Energy Cyclotron Centre, Homi Bhabha National Institute, Kolkata, India\\
$^{133}$ Warsaw University of Technology, Warsaw, Poland\\
$^{134}$ Wayne State University, Detroit, Michigan, United States\\
$^{135}$ Westf\"{a}lische Wilhelms-Universit\"{a}t M\"{u}nster, Institut f\"{u}r Kernphysik, M\"{u}nster, Germany\\
$^{136}$ Wigner Research Centre for Physics, Budapest, Hungary\\
$^{137}$ Yale University, New Haven, Connecticut, United States\\
$^{138}$ Yonsei University, Seoul, Republic of Korea\\
$^{139}$  Zentrum  f\"{u}r Technologie und Transfer (ZTT), Worms, Germany\\
$^{140}$ Affiliated with an institute covered by a cooperation agreement with CERN\\
$^{141}$ Affiliated with an international laboratory covered by a cooperation agreement with CERN.\\

\end{flushleft} 
  %%%%%%% done by webmaster team                                                                            
\end{document}